\documentclass{iaus}
\usepackage{graphicx}

\title[IAUS264~~A comparison of measured and simulated solar network contrast] 
{A comparison of measured and simulated solar network contrast}

\author[N. Afram \& Y.C. Unruh \& S.K. Solanki \& M. Sch\"ussler \& S.K. Mathew]   
{N. Afram$^1$
 \and Y.C. Unruh$^1$
 \and S.K. Solanki$^2$ 
 \and M. Sch\"ussler$^2$ 
 \and S.K. Mathew$^3$}

\affiliation{$^1$Imperial College, London, UK  \\[\affilskip]
$^2$Max-Planck-Institut f\"ur Sonnensystemforschung, D-37191 Katlenburg-Lindau, Germany\\[\affilskip]
$^3$Udaipur Solar Observatory (USO), Rajasthan, India}

\pubyear{2010}
\volume{264}  
\pagerange{}
\jname{IAUS264}
\editors{A.C. Editor, B.D. Editor \& C.E. Editor, eds.}
\begin{document}

\maketitle

\begin{abstract}
Long-term trends in the solar spectral irradiance are important to determine the impact on Earth's climate. These long-term changes are thought to be caused mainly by changes in the surface area covered by small-scale magnetic elements. The direct measurement of the  contrast to determine the impact of these small-scale magnetic elements is, however, limited to a few wavelengths, and is, even for space instruments,  affected by scattered light and instrument defocus. In this work we calculate emergent intensities from 3-D simulations of solar magneto-convection and validate the outcome by comparing with observations from Hinode/SOT. In this manner we aim  to construct the contrast at wavelengths ranging from the NUV to the FIR.
\end{abstract}

\firstsection 

\section{Introduction}
Variations in the long-term spectral solar irradiance, i.e. the changes in the Sun's brightness at a certain wavelength, are significant for their effects on Earth's atmospheric temperature and composition. Magnetic fields concentrated in small structures  influence these long-term changes, with increasing relevance towards the solar limb. Here, we compare observed and simulated  intensity contrasts and investigate their behaviour from disk centre to the limb in 3D MHD simulations with different average vertical magnetic fields.
\section{Simulations and observations}
In this work, we use  3D MHD simulations that have been run with  non-grey radiative transfer using the  MURaM  code (V\"ogler et al. 2005), which  enables realistic simulations of solar magneto-convection  in the photosphere and the upper  layers of the convection zone. The simulations considered here had an average vertical field of 0G, 50G, 200G. The simulations run on a 288 $\times$ 288 surface grid with 100 depth points, corresponding to a solar surface area of  6 $\times$ 6 Mm and a depth of  1.4 Mm.  The intensities are calculated using the spectral synthesis code ATLAS9 (Kurucz 1993) on the simulated model atmospheres.
We determine the root-mean-square (rms) contrast (standard  deviation divided by the mean value) to characterise intensity variations between dark and bright features.

The data used in this work were recorded with the broad-band filter imager mounted on the Solar Optical Telescope (SOT, Tsuneta et al. 2008, Ichimoto et al. 2008, Suematsu etal. 2008, Shimizu et al. 2008) onboard Hinode
       on 8th Nov. 2006 (during the Mercury transit), and were also used to obtain the point-spread-function (PSF) 
       (see Mathew et al. 2009).  The pixel resolution is ~0.05 arsec. The dark correction and flat fielding were done with SSW-IDL routines.
We have calculated intensity (rms) contrasts  for the Hinode images at  different wavelengths (388nm (CN), 430.5nm (G-band), 450nm (blue continuum), 555nm, 668nm) and compared them with contrasts for the MURaM/Atlas images (which were convolved with the corresponding PSF). Figure 1 shows the comparison of simulated (original and convolved) and observed images and their contrast histograms for the blue continuum, where we obtain an rms contrast of $12$\% for the convolved MURaM simulation as well as for the HINODE observations, while the rms contrast for the original simulation is $21.6$\%. Similar results are obtained for the other wavelengths.

\begin{figure*}
\centering
\begin{picture}(250,250)
\put(-80,0){\begin{picture}(0,0) \includegraphics{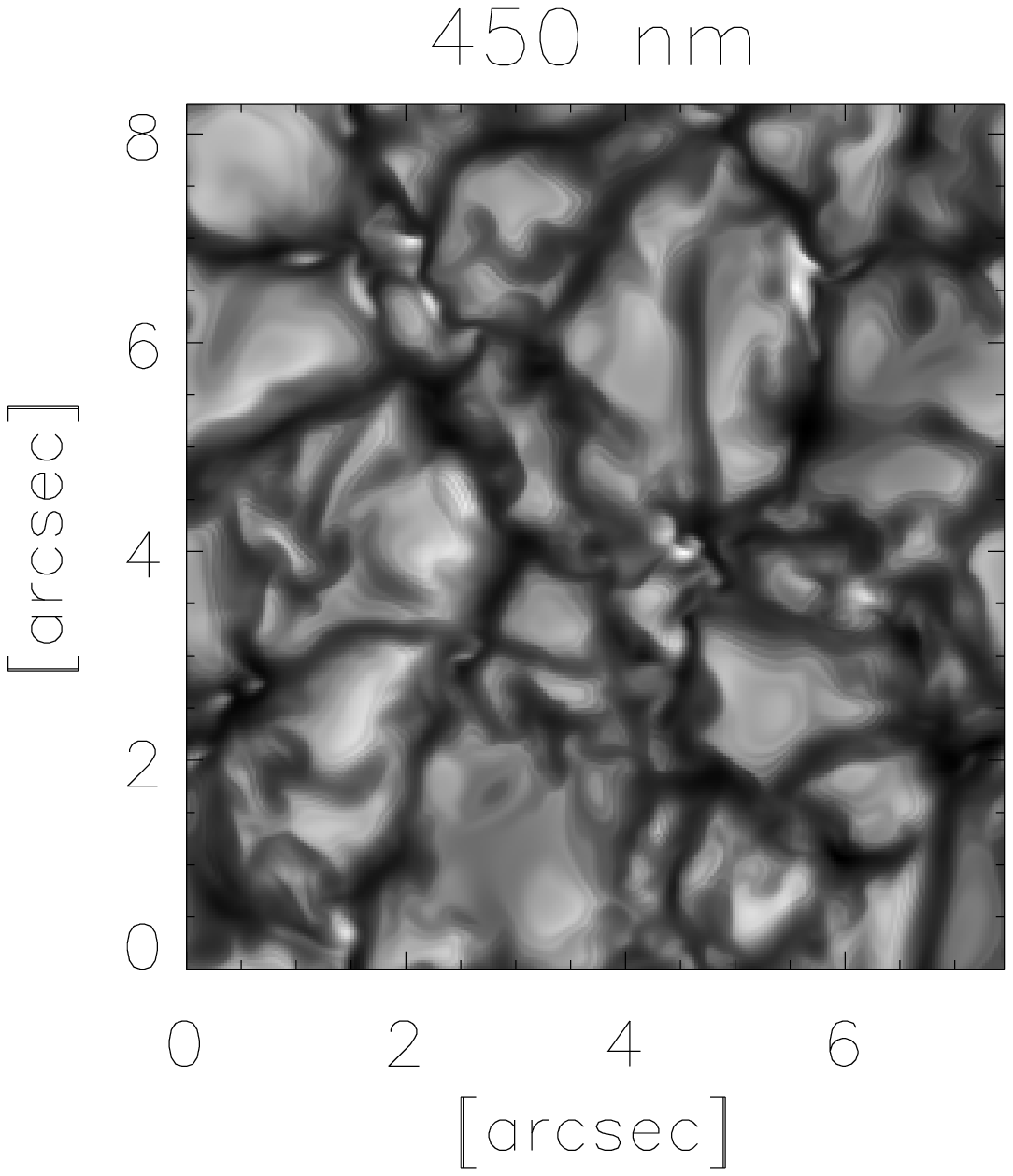} \end{picture}}
\put(65,0){\begin{picture}(0,0) \includegraphics{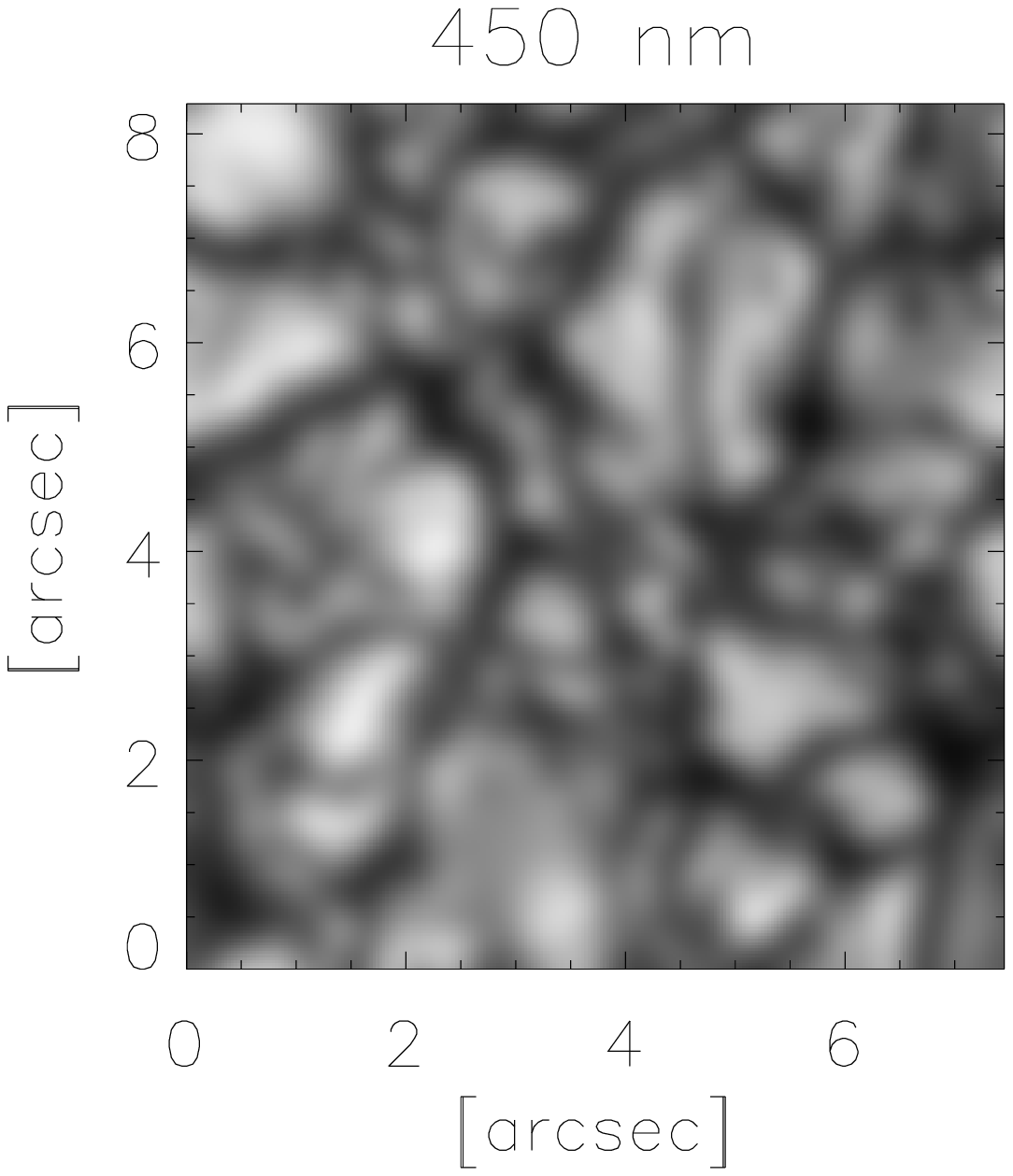} \end{picture}}
\put(180,0){\begin{picture}(0,0) \includegraphics{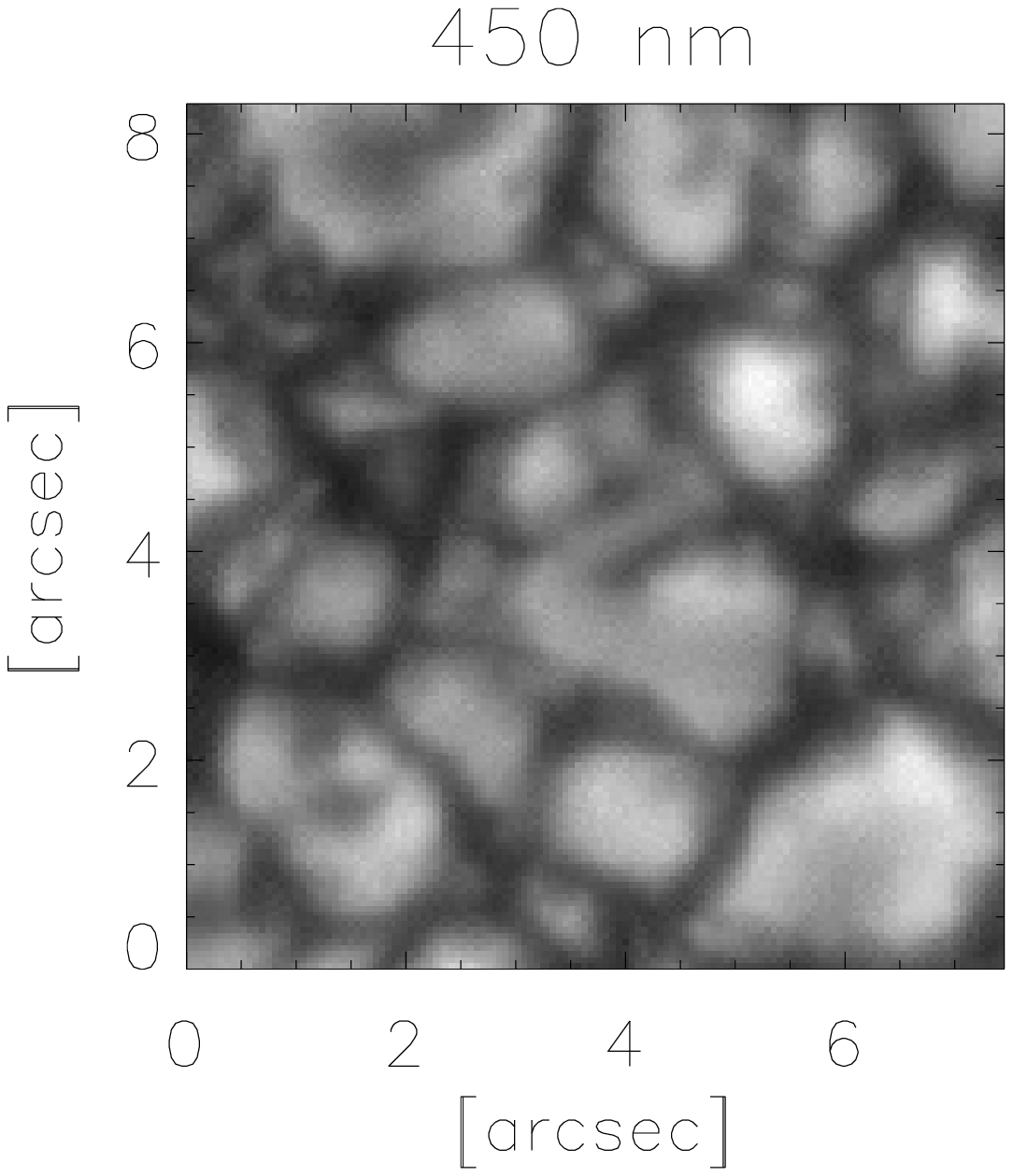} \end{picture}}
\put(-90,-130){\begin{picture}(0,0) \includegraphics{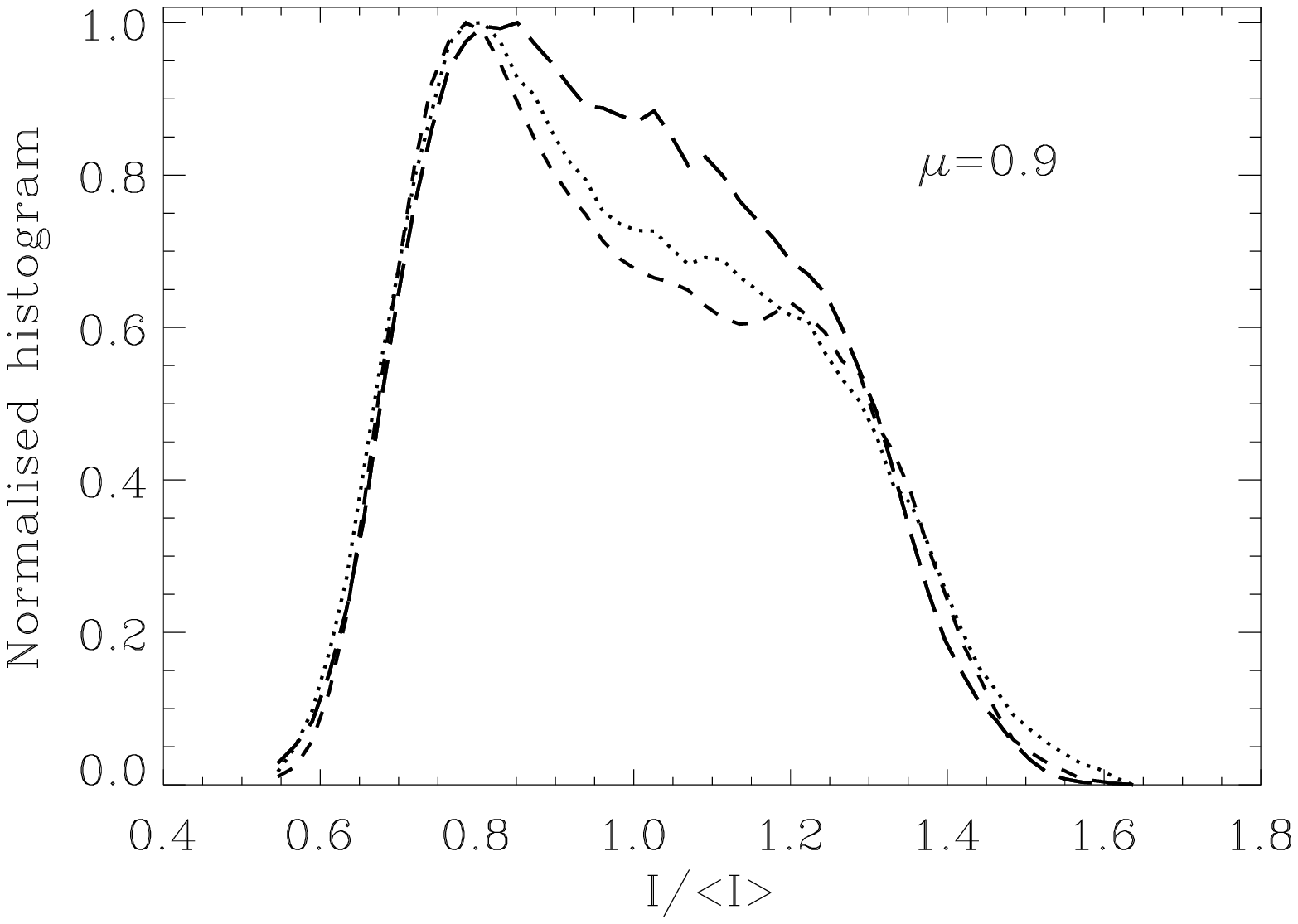} \end{picture}}
\put(122,-130){\begin{picture}(0,0) \includegraphics{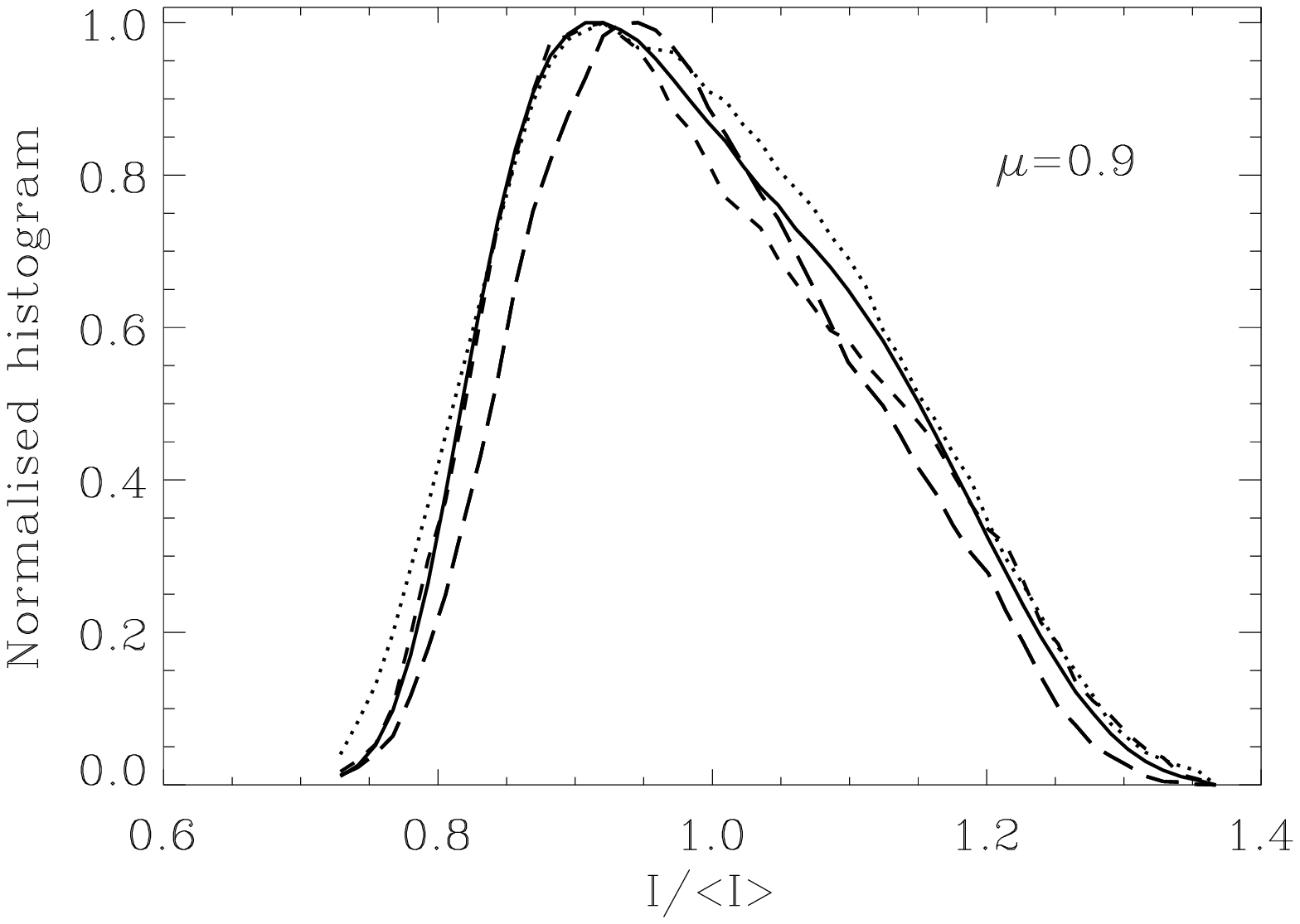} \end{picture}}  
\end{picture}
\caption{Simulations and observations in the blue continuum filtergram. \textit{Top left:} Original simulation snapshot with an average vertical magnetic field of 50G. \textit{Top middle:} Convolved simulation snapshot with an average vertical magnetic field of 50G. \textit{Top right:} Extract from observed image. \textit{Bottom left:} Normalised histograms for the intensity distribution in the original simulations: 0G (dotted), 50G (dashed), 200G (long dashes). \textit{Bottom right:} Normalised histograms for the intensity distribution in convolved simulations and in observations:  0G (dotted), 50G (dashed), 200G (long dashes), observation (thick solid).}
\label{fig:muram50G_hinode_unconv_conv_bc}
\end{figure*}

\section{Centre to limb variation in simulations}

A spatial element of the solar atmosphere radiates differently if it is located at disk centre or nearer to the solar limb, and this spatial evolution of the irradiance behaves differently for different wavelengths. In the UV  magnetic features are comparatively brighter. These changes  become even stronger near the limb and are therefore crucial for spectral solar irradiance variations. 
In Figure 2,  we study the simulated contrasts (i.e. the mean intensity difference between the magnetic and quiet Sun simulation compared to the quiet Sun) as  a function of heliocentric position in five different wavelengths (corresponding to the Hinode/SOT filters). We find that the contrast increases towards the limb and is higher for shorter wavelengths. For the simulation with the higher average vertical field of 200G, the contrasts are significantly larger. The flattening and turnover (e.g. for 388 nm and 430 nm) agree better with measurements of facular contrasts than 1D calculations (e.g. Unruh et al 1999, Ortiz et al. 2002).

  \begin{figure*}
\centering
\begin{picture}(150,150)
\put(-150,-150){\begin{picture}(0,0) \includegraphics{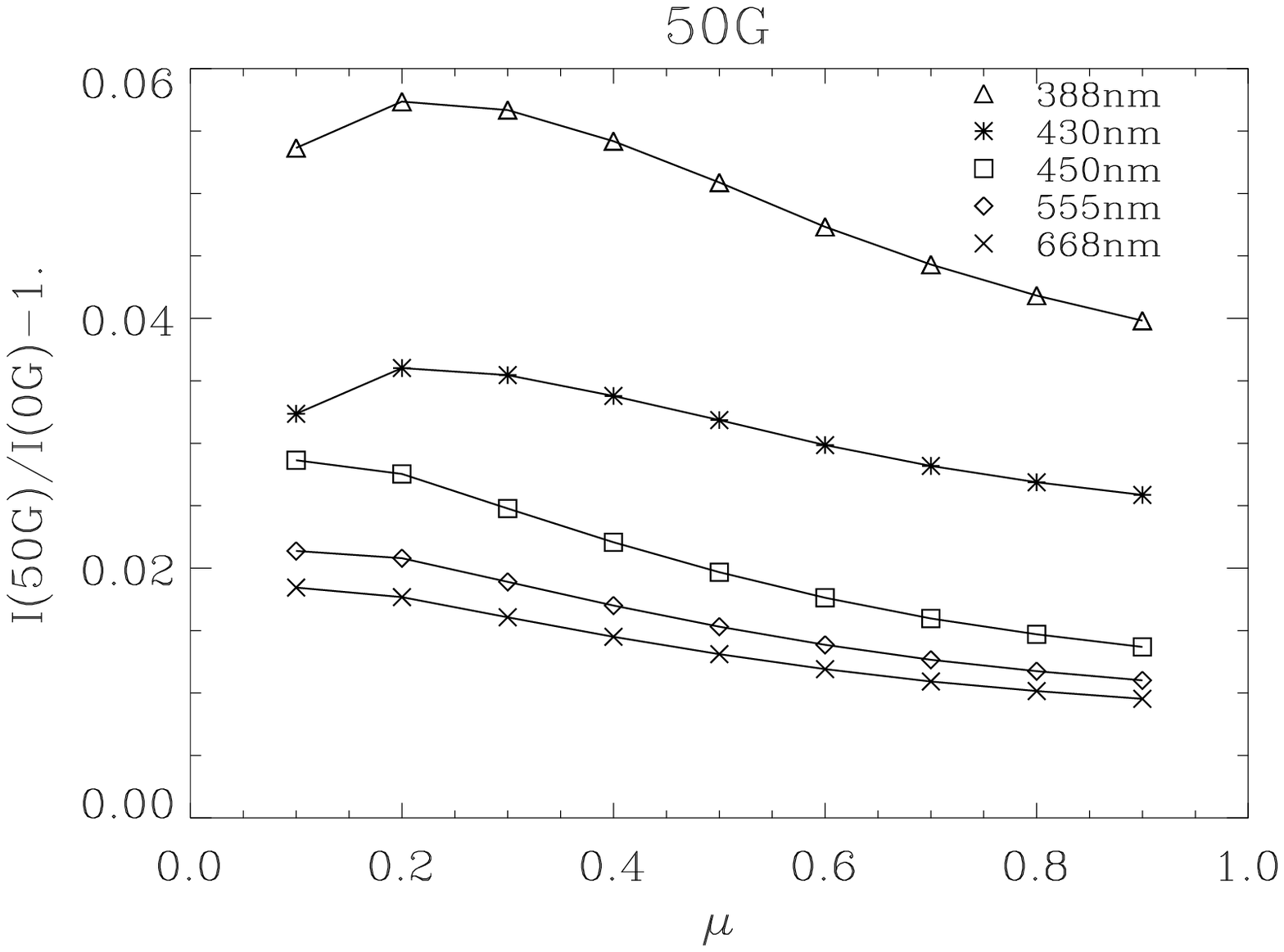} \end{picture}}
\put(50,-150){\begin{picture}(0,0) \includegraphics{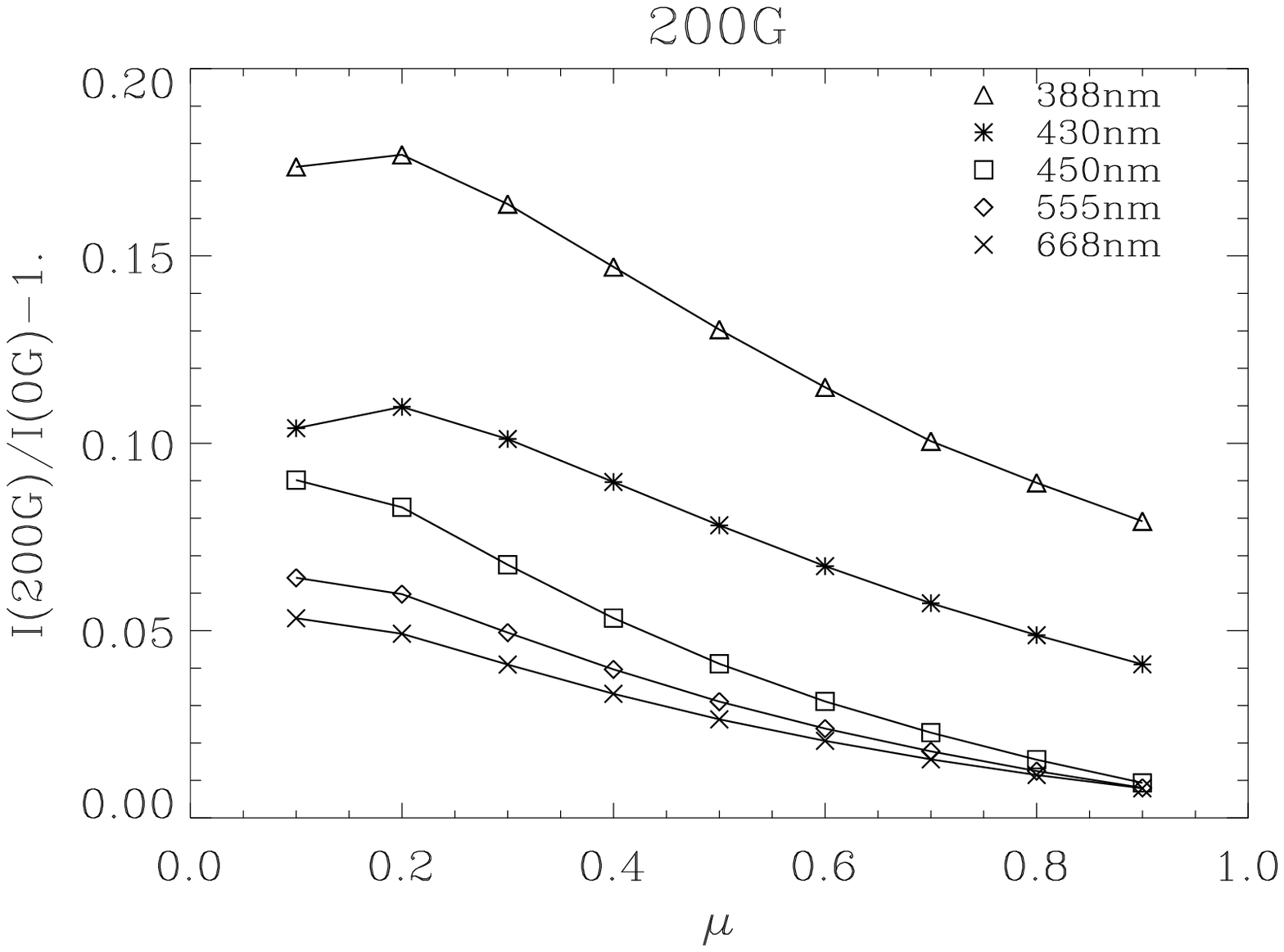} \end{picture}}
\end{picture}
\caption{Contrast as a function of limb angle ($\mu$ $=$ $cos \theta$) for the 5 different Hinode/SOT wavelengths for an average vertical magnetic field of 50G and 200G, respectively. The contrast is given as the mean intensity difference between the magnetic and quiet Sun simulation compared to the quiet Sun.}
\label{fig:clv_50G_200G}
\end{figure*}

\section{Conclusions}
 This analysis will allow us to determine the network and facular contrast as a function of magnetic flux, limb angle, and wavelength. We will then be able to remove the single free parameter in the SATIRE model (e.g. Krivova et al. 2003) and help improve irradiance reconstructions.

\end{document}